\documentclass[12pt]{article}
\usepackage{a4wide,epsfig,psfrag,amsmath,amssymb,cite,scalefnt}
\usepackage{color}
\usepackage{url}

\graphicspath{ {figs/} }

\newcommand{\dd}{{\rm d}}
\newcommand{\nn}{\nonumber}

\newcommand{\ep}{\epsilon}

\allowdisplaybreaks

\begin{document}

\title{\vskip-3cm{\baselineskip14pt
    \begin{flushleft}
    \end{flushleft}} \vskip1.5cm 
Solving differential equations for Feynman integrals by expansions near singular points 
}

\author{
Roman N. Lee$^{a}$,
Alexander V. Smirnov$^{b}$,
\\
Vladimir A. Smirnov$^{c}$ 
\\[1em]
{\small\it (a) Budker Institute of Nuclear Physics,}\\
{\small\it 630090 Novosibirsk, Russia}
\\
{\small\it (b) Research Computing Center, Moscow State University}\\
{\small\it 119991, Moscow, Russia}
\\  
{\small\it (c) Skobeltsyn Institute of Nuclear Physics of Moscow State University}\\
{\small\it 119991, Moscow, Russia}
}
  
\date{}

\maketitle

\thispagestyle{empty}

\begin{abstract}
We describe a strategy to solve differential equations for Feynman integrals by powers series expansions near singular points and to obtain high precision results for the corresponding master integrals. 
We consider Feynman integrals with two scales, i.e. nontrivially depending on one variable.
The corresponding algorithm is oriented at situations where canonical form of the differential equations is impossible.
We provide a computer implementation of our algorithm in a simple example of four-loop generalized sun-set integrals with three equal non-zero masses.
Our code provides values of the master integrals at any given point on the real axis with a required accuracy and a given order of expansion in the 
regularization parameter $\epsilon$.
 
  \medskip

  \noindent

\end{abstract}

\thispagestyle{empty}


\newpage


\section{Introduction}

Evaluating Feynman integrals with differential equations (DE) initiated in 
\cite{Kotikov:1990kg,Kotikov:1991pm} and formulated as a method to evaluate master integrals
in \cite{Remiddi:1997ny,Gehrmann:1999as,Gehrmann:2000zt,Gehrmann:2001ck}
became one of the most powerful methods already much time ago. 
Still this method is under development.
In \cite{Henn:2013pwa} it was suggested to turn from the basis of primary master
integrals 
(i.e. revealed when solving integration by parts relations \cite{Chetyrkin:1981qh})
to the so-called {\em canonical} basis for which the right-hand side of the of system of DE is just proportional to $\epsilon=(4-D)/2$ and the singularities of the matrix 
on the right-hand side of DE are Fuchsian, i.e. have only simple poles in all the singular points of DE.
The first algorithm to arrive at the canonical form was constructed in the case of one variable in Ref.~\cite{Lee:2014ioa} (Such form of DE was called  $\epsilon$-form there).
Besides the private implementation of this algorithm by its author and
several other private implementations, two public implementations, {\tt Fuchsia}~\cite{Gituliar:2016vfa,Gituliar:2017vzm} and {\tt epsilon}~\cite{Prausa:2017ltv}, of the algorithm of Ref.~\cite{Lee:2014ioa} are now available.\footnote{ See also~\cite{Meyer:2016slj,Meyer:2017joq} where an algorithm in the case of two and more variables is described and implemented.}

Once DE for master integrals are converted into an $\epsilon$-form, i.e. one finds an appropriate linear transformation to a canonical basis, solving DE becomes straightforward, order-by-order in $\epsilon$. 
Typically, the corresponding results are expressed naturally in terms of harmonic polylogarithms~\cite{Remiddi:1999ew} or multiple polylogarithms~\cite{Goncharov:1998kja}.
These functions are very well studied. For harmonic polylogarithms, one can apply the package {\tt HPL} \cite{Maitre:2005uu} which encodes various analytical properties and provides the possibility of numerical evaluation with a very high precision. 
For multiple polylogarithms, one can use the computer implementation \cite{GiNaC_code} of the algorithm {\tt GiNaC} \cite{GiNaC} to obtain high-precision numerical values, up to several thousand digits and more.
  
It is well known that the $\epsilon$-form of DE for a given set of the master integrals is not always achievable by rational transformations. 
For massive internal lines it is often required to consider also transformations involving square roots. 
However, even using transformations from this extended class it is not always possible to obtain 
an $\epsilon$-form.\footnote{Recently, a strict criterion of the existence of an $\epsilon$-form was presented in Ref. \cite{Lee:2017oca}.}
The simplest example where an $\epsilon$-form is impossible is given by the two-loop propagator sunset diagram with three identical masses.
In this example, as well in other known examples without $\epsilon$-form, DE can still be reduced
to the form where the right-hand side of the differential system is a linear function of $\epsilon$. 
 
However, 'integrating out' the constant term in such a form of DE appears to be an essentially more complicated 
problem. This can be seen in the known examples where results are expressed in terms of elliptic functions.
In practice, it can happen that such `elliptic' master integrals appear only in a small number of sectors. 
(A sector is specified by a distribution of the set of indices (powers of propagators) into positive and non-positive values.) 
A first example of a calculation of a full set of the master integrals with `elliptic sectors' can be found in  
Ref.~\cite{Bonciani:2016qxi}, where elliptic functions appear only in two sectors and final results are expressed either 
in terms of multiple polylogarithms or, for the elliptic sectors, in terms of two and three-fold iterated integrals suitable 
for numerical evaluation.
Moreover, in Refs.~\cite{Primo:2016ebd,Primo:2017ipr} a strategy to obtain parametric representations 
for master integrals applicable also in situations without $\epsilon$-form 
was described and illustrated through one-, two- and three-loop examples.

Other examples of calculations of individual Feynman integrals in situations where $\epsilon$-form is impossible 
can be found in~\cite{Adams:2016xah,Adams:2017ejb} (see also references therein), where results are expressed in terms of
elliptic generalizations of polylogarithms~\cite{Adams:2016xah} or iterated integrals of modular forms\cite{Adams:2017ejb}. 
One more class of elliptic generalization of multiple polylogarithms was recently introduced in Ref.~\cite{Remiddi:2017har}.
In particular, it includes functions appearing in the $\epsilon$-expansion of the imaginary part of the two-loop massive sunset 
diagram.
However, these new functions do not have the same status as harmonic polylogarithms and multiple polylogarithms, 
at least in the practical sense, i.e. there are no codes to evaluate them at a given point with a desired precision.
Anyway, it looks like we are very far, even in lower loops orders, from answering the following question: `What is the class of functions
which can appear in results for Feynman integrals in situations where $\epsilon$-form is impossible'?

On the other hand, thinking positively, we may say that knowing a differential system and 
the corresponding boundary conditions gives almost as much information about Feynman integrals as knowing their explicit expressions 
in terms of some class of functions. 
In fact, some properties of the integrals are even more accessible via DE. 
In particular, singularities of DE provide a way to examine the branching properties of integrals. 
Numerical values of the integrals can be obtained from a numerical 
solution\footnote{Examples of solving DE for Feynman integrals numerically can 
be found in Refs.~\cite{Czakon:2008zk,Baernreuther:2013caa}.} 
of the differential system.
Many computer algebra systems contain tools to solve this task (e.g. \texttt{NDSolve} procedure in \textit{Mathematica} system). 
However, there is one complication that does not allow to use these tools immediately. 
Namely, we would like to keep $\epsilon$ as a variable and evaluate solutions of DE
as series expansions in $\epsilon$.

The goal of the present paper is to describe an algorithm which enables one to find a solution of a given differential system 
in the form of an $\epsilon$-expansion series with numerical coefficients. 
We describe such an algorithm in the case of Feynman integrals depending on one variable, i.e. with two scales 
where the variable is introduced as the ratio of these scales.
As a  proof of concept, we provide a computer code where this algorithm is implemented for a simple example 
of a family of Feynman integrals where the $\epsilon$-form is impossible.
The general idea behind our approach is to use generalized power series expansions near the singular points of the differential system
and solve difference equations for the corresponding coefficients in these expansions.
This idea is very well known in mathematics. In high-energy physics, its application to Feynman integrals can be found, for example,
in Ref.~\cite{Kniehl:2017ikj}, where three-loop massive vacuum diagrams were 
evaluated.\footnote{Another example, where the general theory of DE was applied for evaluating
expansion of two-scale integrals at a given singular point, can be found in Ref.~\cite{Mueller:2015lrx}.
For this purpose, one can apply various mathematical prescriptions from the theory of DE -- see, e.g.
Ref.~\cite{Henn:2016kjz}, where an algorithm~\cite{Wasow} to obtain first terms of expansion near a singular
point was applied. An approach similar to Ref.~\cite{Mueller:2015lrx} was applied
in Ref.~\cite{Melnikov:2016qoc} to evaluate expansions of solutions of DE  at a given singular point by difference
equations.}

In the next section, we present an algorithm to solve difference equations for coefficients of
the series expansions at a given singular point. In Section~3, we describe a matching procedure which enables
one to connect series expansions at two neighboring points. In Section~4, we describe a computer code based on our algorithm 
and the matching procedure
to evaluate master integrals in a simple four-loop example. Then we conclude with a discussion of perspectives.

\section{Generalized series expansion near a singular point}

Let us have a differential system 
\begin{equation}
\partial_{x}\boldsymbol{J}=M\left(x,\epsilon\right)\boldsymbol{J}\,,
\label{DE}
\end{equation}
where $\boldsymbol{J}$ is a column-vector of $N$ functions, and $M$ is an $N\times N$ matrix with entries 
being rational functions\footnote{Typically, $x$ is the dimensionless ratio of two scales 
for a family of dimensional regularized Feynman integrals depending on two scales.} of $x$ and $\epsilon$. 
Below we will suppress $\epsilon$ in the arguments for brevity.
We assume that all the singular points of the differential system are regular. 
This implies that we can reduce the differential system to a local Fuchsian form in any singular point.  

The general solution of this linear system has the form
\begin{equation}
\boldsymbol{J}\left(x\right)=U\left(x\right)\boldsymbol{C}\,,
\label{DEsolution}
\end{equation}
where $\boldsymbol{C}$ is a column of constants, and $U$ is an evolution operator represented in terms of a path-ordered exponential  
\begin{equation}
U\left(x\right)=P\exp\left[\int dxM\left(x\right)\right]\,.
\label{evolution}
\end{equation}
We want to expand this operator in the vicinity of each singular point.
Without loss of generality, let us consider the expansion near  $x=0$.
It is well known that the expansion has the form
\begin{equation}
U\left(x\right)=\sum_{\lambda\in S}x^{\lambda}\sum_{n=0}^{\infty}\sum_{k=0}^{K_{\lambda}}\frac1{k!}C\left(n+\lambda,k\right)x^{n}\ln^{k}x\,,
\label{evolution_expansion}
\end{equation}
where $S$ is a finite set of powers of the form $\lambda=r \epsilon$ with integer $r$,  $K_\lambda\geqslant0$ is an 
integer number corresponding to the the maximal power of the logarithm. 
We have introduced the factor $1/k!$ for convenience. Our goal is to determine $S$, $K_\lambda$, 
and the matrix coefficients $C\left(n+\lambda,k\right)$. As to the latter, we are going to determine them via recurrence 
relations equipped with initial conditions.

Since we assume that the differential system has only regular singular points, we can reduce it at $x=0$ to normalized Fuchsian form \cite{Lee:2017oca} by means of rational transformations. For the sake of presentation, we will assume that the system is in global normalized Fuchsian form, i.e.,
\begin{equation}
M\left(x\right) = \frac{A_0}{x}+\sum_{k=1}^{s}\frac{A_k}{x-x_k} 
\label{NFF}
\end{equation}
and for any $k=0,\ldots,s$ the matrix $A_k$ is free of resonances, i.e. the difference of any two of its distinct eigenvalues is not integer. 
Note that the $\epsilon$-form is only one example of normalized Fuchsian form, so we allow for a much wider class of 
differential systems which seems to be sufficient for any applications in multiloop calculations. 
In particular, the `elliptic' cases, as a rule, can easily be reduced to a global normalized Fuchsian form. 
Besides, it is easy to generalize our algorithm properly if needed.

In order to obtain a recurrence relation of a finite order, we will first multiply both sides of Eq.~\eqref{DE} 
by the common denominator $xQ(x)$, where 
\begin{equation}
Q\left(x\right)=\prod_{k}\left(x-x_{k}\right)=\sum_{m=0}^s q_{m}x^{m}\,.
\end{equation}
By construction we have $q_0\neq0$. We will also define the polynomial matrix $B\left(x,\alpha\right)$ and its coefficients $B_{m}\left(\alpha\right)$ by 
\begin{equation}
B\left(x,\alpha\right)=Q\left(x\right)\left(xM\left(x\right)-\alpha\right)=\sum_{m=0}^NB_{m}\left(\alpha\right)x^{m}\,.
\end{equation}
Note that $B_{0}\left(\alpha\right)=q_0(A_0-\alpha)$.

Then the recurrence relations read
\begin{multline}
-\mathrm{BJF}(B_{0}(\lambda+n),-q_{0},K_\lambda)C\left(\lambda+n,0..K_\lambda\right)\\
=\sum_{m=1}^s\mathrm{BJF}(B_{m}\left(\lambda+n-m\right),-q_{m},K_\lambda)C\left(\lambda+n-m,0..K_\lambda\right)\,.
\label{recurrence}
\end{multline}
Here
$C\left(\alpha,0..K\right)=\begin{bmatrix}C\left(\alpha,0\right)\\
\vdots\\
C\left(\alpha,K\right)
\end{bmatrix}$ denotes a $(K+1)N\times N$ matrix built from blocks $C\left(\alpha,k\right)$ and the three-letter notation $\mathrm{BJF}$ stands for `Block Jordan Form' defined as
\[
\mathrm{BJF}(A,B,K)=
\underbrace{\begin{bmatrix}A & B & 0 & 0\\
0 & \ddots & \ddots & 0\\
0 & 0 & \ddots & B\\
0 & 0 & 0 & A
\end{bmatrix}}_{K+1}\,.
\] 

Now note that the operator $U$, Eq. \eqref{evolution}, is determined up to a multiplication by a constant matrix from the right. 
We fix it by the condition
\begin{equation}
U(x)\stackrel{x\to 0}\sim x^{A_0}\,.
\end{equation}
This condition is, strictly speaking, mathematically incorrect when the distance between some eigenvalues of $A_0$ is larger 
than one, but it should be understood as the constraint on the leading terms of the expansion for each distinct eigenvalue. 
This condition gives us a way to determine $S$, i.e. the set of distinct eigenvalues of $A_0$, and $K_\lambda$, 
i.e. the highest power of the logarithm in front of $x^\lambda$ in $x^{A_0}$ for each $\lambda\in S$, 
and the leading coefficients $C(\lambda,k)$. We simply determine these parameters by representing
\begin{equation}\label{initial_conditions}
x^{A_0} =\sum_{\lambda\in S} x^\lambda \sum_{k=0}^{K_\lambda} \frac1{k!}C(\lambda,k)\ln^kx\,.
\end{equation}
Now note that the matrix $-\mathrm{BJF}(B_{0}(\lambda+n),-q_{0},K_\lambda)$ on the left-hand side of Eq. \eqref{recurrence} is invertible for $\lambda\in S$ and $n>0$. Indeed, 
\begin{equation}
	\det \mathrm{BJF}(B_{0}(\lambda+n),-q_{0},K_\lambda)= \left(\det B_{0}(\lambda+n)\right)^{K_\lambda+1}=
	q_0^{(K_\lambda+1)n}\left[\det (A_0-\lambda-n)\right]^{K_\lambda+1}
\end{equation}
and both $q_0\neq 0$ and $\det (A_0-\lambda-n)\neq 0$, the latter is due to the absence of resonances in $A_0$ (since if $\det =0$ both $\lambda$ and $\lambda+n$ would be the eigenvalues of $A_0$). Therefore, we can rewrite recurrence relations \eqref{recurrence} as
\begin{equation}
C\left(\lambda+n,0..K_\lambda\right)
=\sum_{m=1}^sT(\lambda,n,m)C\left(\lambda+n-m,0..K_\lambda\right)\,,
\label{recurrence1}
\end{equation}
where 
\begin{multline}
T(\lambda,n,m)=-\left[\mathrm{BJF}(B_{0}(\lambda+n),-q_{0},K_\lambda)\right]^{-1}\mathrm{BJF}(B_{m}\left(\lambda+n-m\right),-q_{m},K_\lambda)\,.
\label{Tcoefs}
\end{multline}
and use \eqref{recurrence1} together with the initial conditions determined\footnote{One also puts $C(\lambda+n,k)=0$ for $\lambda\in S$ and $n<0$.} by Eq. \eqref{initial_conditions}
 in order to construct the generalized power series expansion \eqref{evolution_expansion}. 
Note that the finite-order recurrence relation results in a linear growth of the computational complexity with the number of expansion terms. 

To summarize, the data necessary to obtain the expansion \eqref{evolution_expansion}
are as follows:
\begin{enumerate}
\item The set $S=\left\{ \lambda_{1},\lambda_{2},\ldots\right\} $ of the eigenvalues of the matrix residue $A_0$.
\item For each $\lambda\in S$:
\begin{enumerate}
\item the maximal power of the logarithm $K_{\lambda}$ and the leading coefficients
$C\left(\lambda,0..K_\lambda\right)$ defined by \eqref{initial_conditions}. 
To use the recurrence formula one has to take into account that $C\left(\lambda+n,k\right)=0$
for $n<0$.
\item the matrix coefficients $T\left(\lambda,n,1\right),\ldots,T\left(\lambda,n,s\right)$
which are $\left(K_{\lambda}+1\right)N\times\left(K_{\lambda}+1\right)N$ matrices,
where the dependence on $n$ is explicit.
\end{enumerate}
\end{enumerate}

\section{Matching}

The above considerations enable one to evaluate the evolution operator \eqref{evolution} within the convergence region of the power series \eqref{evolution_expansion}. 
In order to perform an analytical continuation to the whole complex plane, one may use the same approach for the expansion around other singular points. 
Suppose that the next singular point closest to the origin is $x=1$. 
We can construct the evolution operator (\ref{evolution}) also in an expansion near this point.
\begin{equation}
\tilde{U}\left(x\right)=P\exp\left[\int dxM\left(x\right)\right]\,.
\label{evolution1}
\end{equation}
In general, due to the above mentioned freedom in definition of the evolution operator we have 
\[
U\left(x\right)=\tilde{U}\left(x\right)L \,.
\]
where $L$ is some constant matrix. If the convergence regions of the power series in $U$ and $\tilde U$ overlap, we may fix $L$ by picking some point in the intersection of these regions. E.g. at $x=1/2$ we have\footnote{The convergence radius of the power series is equal to the distance to the closest singularity, so $x=1/2$ necessarily belongs to the convergence region of the series representation of $U$. We also assume here that it belongs to the convergence region of $\tilde U$.} $L=\tilde{U}^{-1}\left(1/2\right)U\left(1/2\right)$,
i.e., finally, in the whole convergence region of $\tilde U$ we have
\[
U\left(x\right)=\tilde{U}\left(x\right)\tilde{U}^{-1}\left(1/2\right)U\left(1/2\right)\,.
\]
Acting in the same way, we may, in principle, extend the definition of $U$ onto the whole complex plane of $x$. 
In fact, this is a general approach to the analytical continuation of a function defined by a converging power series. 
In order to reach an arbitrary finite point of the complex plane, we are likely to need also expansions near the regular points 
(reducible to the considered case by putting $A_0=0$) and/or M\"obius transformations of the variable. 
In the case where the singularities lie on the real axis and if we are interested in the evaluation of Feynman integrals for real $x$, 
we can avoid expansions near regular points and rely only on the M\"obius transformations. 
Suppose, e.g., that we have the following sequence of the singular points
\begin{equation}
	x_0<x_1<\ldots x_s<\infty=x_{s+1}=x_{-1}\,.
\end{equation}
Then for each $0\leqslant k\leqslant s$ we make the variable change
\begin{equation}
y_k(x)=\frac{a x+b}{c x+d}
\end{equation}
which maps the points $x_{k-1},\ x_k,\ x_{k+1}$ to $\mp 1,\ 0,\ \pm 1$, respectively.\footnote{Explicitly we have $y_k(x)=\pm \frac{\left(x-x_k\right) \left(x_{k+1}-x_{k-1}\right)}{(x-x_{k+1})(x_{k-1}-x_k)+(x-x_{k-1})(x_{k+1}-x_k)}$.
}
It is convenient to choose the sign in such a way that the cuts of the non-integer powers and logarithms appearing in the series expansions coincide with the cuts of the integral.

\section{\label{sec::code}Implementation}
 
\begin{figure}[h] 
\begin{center}
\includegraphics[width=0.25\textwidth]{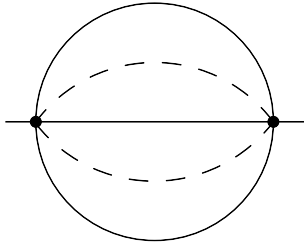}
\caption{\label{fig::ss00mmm}
The generalized sunset graph with two massless and three massive lines with the same mass.}
\end{center}
\end{figure}

The four master integrals we evaluate form a basis of the following 
family of integrals:
\begin{eqnarray}
F_{a_1,\ldots,a_{14}}&=&\int\ldots\int\frac{
\dd^D k_1\ldots\dd^D k_4\;(k_1 \cdot p)^{a_6} (k_2\cdot p)^{a_7} (k_3\cdot p)^{a_8} (k_4\cdot p)^{a_9} }
{(-k_1^2)^{a_1}(-k_2^2)^{a_2}(m^2-k_3^2)^{a_3}(m^2-k_4^2)^{a_4}(m^2-(\sum k_i+p)^2)^{a_5}}
\nn  \\ \hspace*{-80mm} && \times
(k_1\cdot k_2)^{a_{10}} 
(k_1\cdot k_3)^{a_{11}} (k_1\cdot k_4)^{a_{12}} (k_2\cdot k_3)^{a_{13}} (k_2\cdot k_4)^{a_{14}} \;,
\label{fam18ind}
\end{eqnarray}
where $p$ is the external momentum and $m$ is the mass of three lines. They correspond to the generalized
sunset graph shown in Fig.~\ref{fig::ss00mmm}. We introduce $x = p^2/m^2$. 
  
There are four master integrals in this family.
As the primary master integrals we choose the following basis:
\begin{equation}
\boldsymbol{J_0}=\{F_{1,1,1,1,1,0,\ldots,0}, \, F_{1,1,2,1,1,0,\ldots,0}, \, F_{1,2,1,1,1,0,\ldots,0}, \,F_{1,2,1,1,2,0,\ldots,0}\} \,.
\label{MIs}
\end{equation}  
  
We derive DE for $\boldsymbol{J_0}$ in a straightforward way.  When taking derivatives with respect to $x$
one can apply {\tt LiteRed}~\cite{Lee:2012cn,Lee:2013mka} to do this
automatically. The derivatives are then expressed in terms of integrals of the given family.
Solving integration by parts relations with an IBP-reduction code\footnote{In our paper, we
use  {\tt FIRE}~\cite{Smirnov:2008iw,Smirnov:2013dia,Smirnov:2014hma} 
in combination with {\tt LiteRed}~\cite{Lee:2012cn,Lee:2013mka}.}, one expresses these
derivatives as linear combinations of the primary master integrals and obtains a system of linear DE
which has the form (\ref{DE}).

The matrix in the corresponding DE, as well as other entries mentioned in the section,
can be can be downloaded from \url{https://bitbucket.org/feynmanintegrals/dess}.
One uses
\begin{verbatim}
{M, T, Ti, Mf} = << "Data/TransformationData";
\end{verbatim}
Here $\mathtt{M}$ is the matrix in the DE for the basis of the chosen primary master integrals (\ref{MIs}) and
$\mathtt{Ti}$ is $T^{-1}$.
We turn to the basis $\boldsymbol{J}=T^{-1}\cdot \boldsymbol{J_0}$ for which
we have the matrix $\mathtt{M_f}$ with normalized Fuchsian singularities at any singular point
in the corresponding DE (\ref{DE}). We have $M_f=T^{-1}(M\cdot T-\partial_{x}T)$.
We find the new basis with the help of the algorithm of Ref.~\cite{Lee:2014ioa}.
 
To fix boundary conditions we choose the point $x=0$ where the integrals of the given family
become vacuum integrals. To evaluate the four master integrals at $x=0$ we derive onefold 
Mellin-Barnes representations for them and obtain the possibility to achieve a high precision for any given
coefficient in the $\ep$-expansion. We restricted ourselves to the accuracy of 500 digits but one can increase
it to 1000 digits and more.

The singular points are $x_0 = 0, x_1 = 1, x_2 = 9$ and $x_3 = x_{-1} = \infty.$
We solve difference equations for coefficients in series expansions near singular points
according to the algorithm described in Section~2. The corresponding results are encoded in a file present in the package:
\begin{verbatim}
{L, cis, cisrule} = Get["Data/BoundaryConditions"];
\end{verbatim}
Here $\mathtt{L}$ is a constant matrix (see Section~2) and the list $\mathtt{cis}$ defines the required information about the primary masters. The list  has the form $\mathtt{\{ci[1,\epsilon -1,0],{ci}[2,0,0],{ci}[3,\epsilon -1,0],{ci}[4,0,0]\}}$, where $\mathtt{ci[j,n,k]}$ denotes the coefficient in front of $x^n\ln^k x$ in $j$-th primary master. 
The list of replacement rules $\mathtt{cisrule}$ contains this required information which was obtained using different techniques,
in particular, Mellin-Barnes representations.
 
The matching procedure described in the previous section is performed 
in our example as follows. The variable changes corresponding to the singular points are $f_0 = x/(2 - x), f_1 = (x - 1)/(1 + 7x/9), 
f_2 = (9 - x)/(7 + x), f_3= -9/( 2 x - 9)$. For example, the first 
function maps 0 to 0, 1 to 1 and infinity to $-1$.
In new coordinates the radius of convergence is equal to 1, however, the 
convergence is very slow when approaching the boarder of the convergence domain.

For adjacent regions $i$ 
and $i+1$ we search the best possible matching point which is such $x$ 
that it lies between $x_i$ and $x_{i+1}$ and that $|f_i(x)| = |f_{i+1}(x)|$.
In our case we result in matching points $\{-3,3 (3-2 \sqrt 2),3,3 (3+2 \sqrt 2)\}$.

The matching points are separating the singular points.
We have 
\[
-\infty < -3 < 0 < 3 (3-2 \sqrt 2) < 1 < 3 < 9 < 3 (3 + 2  \sqrt 2) < \infty.
\]
Now to obtain the values in a region different from $(-3, 3(3-2 \sqrt 2))$ 
we have to perform matching by moving in the positive or negative direction.
The regions $(3 (3-2 \sqrt 2),3)$  and $(3(3+2 \sqrt 2),-3)$ (around infinity) 
are adjacent and one matching is enough. For the remaining 
$(3,3 (3 + 2 \sqrt 2))$ region one performs two matchings.
This procedure is performed automatically in the code $\mathtt{DESS.m}$ we provide.
The basic function is  
\begin{verbatim}
DESS[rdatas, x, x0, oe, np]
\end{verbatim}
It builds the evolution operator near a given 
point $x_0$, where $\mathtt{oe}$ is the order in $\epsilon$, $\mathtt{np}$ is the required 
precision, and $\mathtt{rdatas}$ contains all the required information about coefficients
in expansions at all the singular points in a 
special format.

The action of this procedure is performed with the help of the following auxiliary functions: 
\begin{verbatim}
FindLFT[x, {k,l,m}] 
\end{verbatim}
finds the M\"obius transformation in $x$ 
which maps ${k, l, m}$ to $(-1,0,1)$;
\begin{verbatim}
FindMPoint[x, {f1, f2}] 
\end{verbatim}
finds the matching point $x$ with 
$|f_1(x)|=|f_2(x)|$ such that $x$ is between $f_1^{-1}(0)$ and $f_2^{-1}(0)$;
\begin{verbatim}
InverseLFT[x, f]
\end{verbatim}
returns the inverse linear transformation function.

The manipulations with series expansion are performed in
the auxiliary basis $\boldsymbol{J}$, rather in the primary basis $\boldsymbol{J_0}$.
For the evaluation of $\boldsymbol{J_0}$, one takes into account the relation between the bases and
applies the command
\begin{verbatim}
(T /. x -> x0).DESS[rdatas, x, x0, oe, np].L.(cis /. cisrule)   
\end{verbatim}
to evaluate the set of the primary master integrals (\ref{MIs}) at the point $x_0$ (different from the singular points)
in an $\epsilon$-expansion up to the order $\mathtt{oe}$ with the accuracy $\mathtt{np}$.
   
To test our code we ran our procedure with $\mathtt{oe}=15$ and $\mathtt{np}=75$ 
at the sample points  ${-10, -3/2, 1/3, 2/3, 2, 4, 12, 25}$ which lie between the singular and matching points
and confirmed our results with the code {\tt FIESTA} \cite{Smirnov:2015mct}.
For example, at $x_0=25$, we obtain the following result (shown with a truncation to 10 digits) for the first primary integral:
\begin{eqnarray}
-\frac{0.25}{\epsilon^4} + \frac{2.125}{\epsilon^3} - \frac{0.2391337000}{\epsilon^2} -  \frac{5.2663306926}{\epsilon}
- 185.9464179437 + 6.5261388472 \,{\rm i}
&& \nn \\  && \hspace*{-155mm}
- (1825.1476432369 - 48.9550593728 \,{\rm i}) \epsilon 
 - (8406.8551978029 -  176.0638485153 \,{\rm i}) \epsilon^2 
 \nn \\  && \hspace*{-140mm} 
 - (58330.4283767260 -  401.9617475893 \,{\rm i}) \epsilon^3  
\,. \nn
\end{eqnarray}

In fact, the maximal order of expansion in $\epsilon$ and the maximal accuracy
is determined by the boundary conditions where expansion of boundary vacuum integrals 
is included up to $\epsilon^3$ with the accuracy of 500 digits. This results in
an $\epsilon$-expansion up to $\epsilon^3$ of our primary master integrals.
However, we recommend to set $\mathtt{oe}=15$ because high negative powers of $\epsilon$
appear in calculations. Moreover, we recommend to add the value 25 to the desired precision 
$\mathtt{np}$, for a similar reason.   
   
One more command of our code is denoted in the same way but has one more argument:   
{\tt DESS[rdatas, x, f(x), oe, np, nt]}.   
It can be used to obtain a required number {\tt nt} of terms of expansion near a 
given singular point, i.e. $x_0 \in \{0,1,9,\infty\}$.
For the three finite singular points,  one can request an expansion in powers of {\tt f(x)}
which can be any function of the form $f(x)=(x-x_0)/(ax+b)$ different from constant, in particular, 
$f(x) $ can be $\pm (x-x_0)$. For example, at $x_0=9$, the choice $f(x)=9-x$ looks natural and 
leads to an expansion of the primary master integrals in powers and logarithms of $9-x$ with real coefficients. For the singular point $x_0 =\infty$,
one can choose $f(x)=(ax+b)^{-1}$, e.g., $\pm1/x$. Here also the choice $-1/x$ is natural for the same reason as above.

The output of this command (with one more argument) is in the form of a set
of replacements ${n,j,k} \to ...$ which give results for the coefficients
$C\left(n+\epsilon j,k\right)$ in Eq.~(\ref{evolution_expansion}) in the expansion of
the evolution operator near a singular point.
A result in the form of Eq.~(\ref{evolution_expansion}) can be obtained from this
result by applying the command
\begin{verbatim}
FromCoefficientRules[..., {x, x^\[Epsilon], Log[x]}]
\end{verbatim}

Similarly to the evaluation of the master integrals at a given non-singular point,
one has to multiply {\tt DESS[rdatas, x, f(x), oe, np, nt]}  by {\tt L.(cis /. cisrule)}
from the right and by an expansion of the transformation matrix $T$ near $x=x_0$
from the left and then reexpand the product of all the factors at $x=x_0$.

The results for the evolution operator given by {\tt DESS[rdatas, x, f(x), oe, np, nt]}
are linear combinations of powers of $(\pm (x-x_0))^{n+j\epsilon}$ so that it is possible
to select contributions for specific $j$ at this level.
For example, one can arrive at results for the naive part of the expansion of
the primary master integrals near a given  finite singular point by selecting only integer powers.
In fact, near $x_0=0$ and $x_0=1$, we have only Taylor expansions of the master integrals in our example.
We have exponents $x^{j\epsilon}$, with j=1,2,3,4, in the expansion at infinity but this does not
mean that there is no naive expansion. The point is that the limit $x\to \infty$
corresponds to the limit, where $m^2\ll |p^2|$, so that the naive expansion in this limit
reduces to the expansion of integrands in Taylor series in $m^2$.
If one is oriented at this very limit it is reasonable to introduce a dimensionless variable in
another way, as $x=m^2/p^2$, and then the naive expansion will be in integer powers
of this variable.

\section{\label{sec::conc}Conclusion}   
   
We have presented an algorithm for the numerical evaluation of a 
set of master integrals depending nontrivially on one variable 
at a given real point with a required accuracy.
The algorithm is oriented at situations where canonical form 
of the DE is impossible. We have provided a computer implementation 
of the algorithm in a simple example.
This code is similar in spirit to the well-known existing codes 
to evaluate harmonic polylogarithms~\cite{Remiddi:1999ew} and 
multiple polylogarithms~\cite{Goncharov:1998kja},
where the problem of evaluation reduces to summing up appropriate series.

We hope that one can use our algorithm and implement it to evaluate 
master integrals in situations where an analytic evaluation is 
problematic.
In fact, we have provided more than the code for the evaluation of 
the four master integrals we considered because our package includes tools 
for a decomposition of the real axis into domains, a subsequent 
mapping and an introduction of appropriate new variables.
We are thinking of a more general package which would include 
an automation of as many steps of the presented algorithm as possible. 
Input data of this package would be a matrix in DE in the normalized Fuchsian form
(defined near Eq.~(\ref{NFF})). Output data would be the evolution operator
in an epsilon expansion up to a required order with a required accuracy.
In addition to the existing tools, the future package needs at least 
an implementation of the algorithm of Section~2 to solve difference equations 
for series expansions at the singular points.
 
Of course, one can hardy construct a general algorithm to fix boundary 
conditions because, usually, the choice of the corresponding point and 
the way to obtain data for the boundary conditions is done in every 
situation in a special way. Still we can suggest a format for including 
information about the boundary conditions for using it in our future package. 
Anyway, our future package would check if a given system of DE is already 
in a global normalized Fuchsian form, with singularities on the real axis, and, if this 
is true, the package would automatically construct the evolution operator 
in an expansion up to a required order.
 
We discussed the problem of evaluation of Feynman integrals with two scales, i.e.
dependent on one variable, $x$. However, one can apply DE even in the case of
one-scale integrals by introducing an extra scale, solving DE with the
respect to the ratio of the two scales, $x$, and then picking a contribution
to the expansion at the point where $x$ tends to its 
original value~\cite{Henn:2013nsa}. 
The second form of the call of $\mathtt{DESS}$ allows one to find the coefficients of the expansion 
of the primary master integrals
near a given singular point $x_0$. Then it is easy to separate the `naive' part of the expansion,
i.e. the contribution of the non-negative integer powers of $x-x_0$
and to find the `naive' values of the primary integrals at $x=x_0$.
For example, for the integrals considered in the previous section, this procedure can provide naive values
at $x=1$, i.e. integrals considered from the scratch with $p^2$ set to $m^2$ which are
nothing but typical integrals appearing in the evaluation of the $g-2$ factor.

\section*{Acknowledgments}
V.S. is grateful to Oleg Veretin for useful discussions.
The work of A.S. and V.S. was supported by RFBR, grant 17-02-00175A.
The work of R. Lee was supported by the grant of the “Basis” foundation for theoretical physics.

\end{document}